\documentclass[pra,twocolumn,superscriptaddress,dvips,floatfix,aps,showpacs,10pt]{revtex4-1}  
\usepackage{graphicx}  % needed for figures
\usepackage{dcolumn}   % needed for some tables
\usepackage{bm}        % for math
\usepackage{amssymb}   % for math
\usepackage{amsmath}
\usepackage{amsfonts}
\usepackage{epsfig}
\usepackage{psfrag}
\usepackage{braket}
\usepackage{color}
\usepackage{textcomp}
\usepackage{latexsym}
\usepackage{hyperref}

\begin{document}

\title{Self trapping in the two-dimensional Bose-Hubbard model}
%\date{\today}

\author{Andrew Jreissaty}
\affiliation{Department of Physics, Georgetown University, Washington, DC 20057, USA}
\affiliation{Department of Physics, The Pennsylvania State University,
University Park, Pennsylvania 16802, USA}
\author{Juan Carrasquilla}
\affiliation{Department of Physics, The Pennsylvania State University,
University Park, Pennsylvania 16802, USA}
\author{Marcos Rigol}
\affiliation{Department of Physics, The Pennsylvania State University,
University Park, Pennsylvania 16802, USA}

\begin{abstract}
We study the expansion of harmonically trapped bosons in a two-dimensional lattice 
after suddenly turning off the confining potential. We show that, in the presence 
of multiple occupancies per lattice site and strong interactions, the system 
exhibits a clear dynamical separation into slowly and rapidly expanding clouds. 
We discuss how this effect can be understood within a simple picture by invoking 
doublons and Bose enhancement. This picture is corroborated by an analysis of the 
momentum distribution function in the regions with slowly and rapidly expanding bosons.
\end{abstract}

\maketitle

In contrast to standard time-of-flight measurements after turning off all trapping
and optical lattice potentials, in which the role of interactions can either be accounted 
for in a straightforward way or neglected, recent theoretical \cite{rigol_muramatsu_04,
rigol_muramatsu_05,minguzzi_gangardt_05,mrigol_muramatsu_05,rodriguez_manmana_06,campo_08,
fabian_rigol_08,buljan_pezer_08,fabian_manmana_09,jukic_pezer_08,hen_rigol_10,
jreissaty_carrasquilla_11,muth_petrosyan_12,langer_schuetz_12,bolech_meisner_12,
iyer_andrei_12,iyer_guan_13,vidmar_langer_2013} and experimental 
\cite{schneider_hacke_12,ronzheimer_schreiber_13} studies have shown 
that expanding interacting particles in low-dimensional geometries and/or in the 
presence of lattice potentials can lead to surprising and interesting effects [see 
Ref.~\cite{cazalilla_citro_review_11} for a review of some of these effects in 
one dimension (1D)]. Among those, and of particular relevance to this work,
have been the observation of emerging (quasi)coherent correlations during the 
expansion of Mott insulators in 1D \cite{rigol_muramatsu_04,mrigol_muramatsu_05,
rodriguez_manmana_06,fabian_rigol_08} and higher-dimensional \cite{hen_rigol_10,
jreissaty_carrasquilla_11} lattices, as well as self-trapping and quantum 
distillation of fermions \cite{fabian_manmana_09} and bosons \cite{muth_petrosyan_12} 
in 1D.

Expansions of fermions \cite{schneider_hacke_12} and bosons \cite{ronzheimer_schreiber_13} 
have also been studied in optical lattices in experiments. 
In the fermionic case, the systems were prepared in a band-insulating state and 
expansion dynamics were studied for different values (and signs) of the on-site 
interaction. Remarkably, it was found that the dynamics transitions from 
ballistic at very weak interactions to bimodal (a slowly expanding spherical core was 
surrounded by a ballistically expanding square-shaped cloud) with increasing interactions. 
Qualitatively similar results were obtained for bosons initially prepared 
in a Fock state (with $n=1$ or $n>1$) that expanded under different strengths 
(and signs) of the on-site interaction. Motivated by those results, we study the 
expansion of bosons with repulsive interactions in the presence of multiple occupancies. 
We show that, within a mean-field approximation, the system dynamically separates 
into slowly and rapidly expanding clouds, similar to what has been seen in experiments. 
Our results emphasize the importance of self-trapping~\cite{fabian_manmana_09,muth_petrosyan_12} 
in the presence of multiple occupancies and strong interactions in two 
dimensions (2D).

We consider 2D lattice bosons modeled by the Bose-Hubbard 
model~\cite{fisher_weichman_89} with an additional harmonic trap
\begin{eqnarray}
\hat{H} = &-&J\sum_{\langle i,j\rangle} \left(\hat{a}_i^\dag \hat{a}_j + 
\text{H.c.}\right) + \frac{U}{2}\sum_i \hat{n}_i(\hat{n}_i -1) \nonumber\\
&+& \sum_i (V_i - \mu) \hat{n}_i, 
\end{eqnarray}
where standard notation has been used \cite{jreissaty_carrasquilla_11}. 
$V_i = V r_i^2$ models the harmonic potential (with strength $V$), and $r_i$ (to 
be given in units of the lattice spacing) is the distance of site $i$ 
with respect to the center of the trap. 

We study the expansion dynamics after turning off the confining potential ($V=0$) at 
time $t=0$. By selecting $U$ to be smaller than the mean-field critical value 
$(U/J)_c \simeq 23.31$ for the formation of the $n=1$ Mott insulator (we select 
$U/J\leq 23$), the initial state is always taken to be in a superfluid phase with 
an on-site density $1\leq n_{\text{center}} \leq 2$ at the center of the trap. Note 
that $U$ and $J$ are not changed when the trap is turned off. The expansion 
dynamics are studied using the time-dependent Gutzwiller mean-field approximation 
\cite{amico_penna_98,jaksch_venturi_02,jreissaty_carrasquilla_11,snoek_11,supmat}. 

\begin{figure}[!t]
\centering
\includegraphics[width=0.35\textwidth]{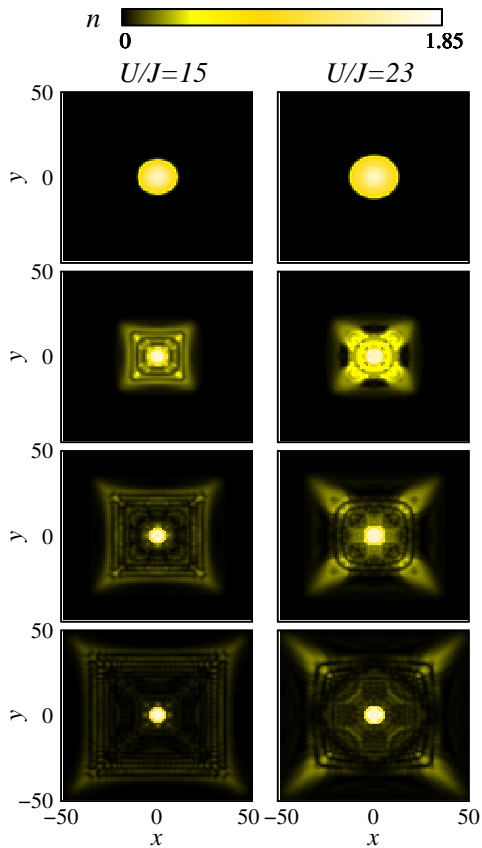}
\caption{(Color online) Density profiles after turning off the trap in systems with 
$U/J=15$ (left) and $U/J=23$ (right). The times depicted are 
$t=0$, $7$, $14$, and $21$ (from top to bottom). For both systems, $V/J=0.161$ and 
$n_{\text{center}} = 1.6$ at $t=0$. Times are reported in units of $\hbar/J$.}
\label{fig:densityProfiles}
\end{figure}

In Fig.~\ref{fig:densityProfiles}, we show results for the density profiles in 
systems with two different values of the on-site interaction [$U/J=15$ (panels) 
and $U/J=23$ ( panels)] and for different times after turning off the trap. 
For the times shown, it can be seen that a significant number of bosons remain 
in the region where $n>1$ at all times (dubbed ``slowly expanding cloud''), while 
the rest of the cloud expands (dubbed ``rapidly expanding cloud''). A comparison 
between the results for $U/J=15$ and $U/J=23$ makes apparent that this effect is 
strengthened when the interaction strength is increased. For strong interactions, such a
self-trapping can be partially understood to be a consequence of energy conservation, 
which forbids the breakup of a double occupancy (doublon) because the resulting excess 
energy cannot be released into the system \cite{winkler_thalhammer_06,fabian_manmana_09}.
Furthermore, pairs of adjacent doublons effectively attract each other, thus increasing 
the stability of the central $n>1$ region. The latter is the result of an energy gain 
due to virtual tunneling transitions between individual bosons belonging to two adjacent 
doublons~\cite{petrosyan_schmidt_07,muth_petrosyan_12,carleo2012}. Within this picture,
a region with dominant doublon occupation can expand only in a time scale
$\propto J^2/U$, due to second order tunneling processes.

On the other hand, bosons that expand into the empty lattice do so in a ballistic 
manner resembling the one observed during the expansion of a Mott insulator with 
$n=1$ \cite{jreissaty_carrasquilla_11}. There, the fastest expansion was seen 
along the diagonals, where the fastest group velocity can develop in this geometry. 
This effect is accentuated as the interaction strength is increased. As mentioned  
above, a similar bimodal expansion in which a nonexpanding core 
was surrounded by ballistically expanding particles (with a square-shaped distribution) 
has been observed in recent experiments that considered different initial states 
for fermions \cite{schneider_hacke_12} and bosons \cite{ronzheimer_schreiber_13}.

Another noticeable characteristic of the slowly expanding cloud 
(Fig.~\ref{fig:densityProfiles}) is that its site occupancies fluctuate in space 
and time, and even reach values that are greater than those in the initial state. 
This is similar to the quantum distillation process observed in fermions in 
one dimension \cite{fabian_manmana_09}, in which doublons grouped together in the 
center of the system. However, in contrast to the latter case,
full distillation (all doublons forming a cluster) does not occur here. 
Figure~\ref{fig:densityProfiles} indicates ($n$ is always smaller than two), 
and an analysis of single occupancies confirms, that a fraction of the bosons 
that remain in the slowly expanding cloud are in singly occupied sites (monomers).

The fact that bosonic monomers can remain in the slowly expanding cloud while fermionic 
ones do not \cite{fabian_manmana_09} is a consequence of Bose enhancement, 
as argued for the one-dimensional case in Ref.~\cite{muth_petrosyan_12}. Extending 
those arguments to two dimensions, this can be understood as follows. In the presence
of strong repulsive interactions, the dispersion relation of bosons within the region 
of density $n\leq 1$ is, to a good approximation, the same as that of hard-core bosons 
in the vacuum $\epsilon_{\mathbf{k},\text{exp}} = -2J(\cos k_x + \cos k_y)$
\cite{hen_rigol_10,jreissaty_carrasquilla_11}. In contrast, within the slowly expanding cloud 
where $n>1$, monomers (again approximated by hard-core bosons) propagate via resonant 
single-particle hopping with a Bose enhanced amplitude $J'=2J$. Thus, the dispersion 
relation of monomers is $\epsilon_{\mathbf{k},\text{trap}} = -4J(\cos k_x + \cos k_y)$. 
This means that in the rapidly expanding cloud bosons have $\epsilon_{\mathbf{k},\text{exp}}\in[-4J,4J]$ 
while in the slowly expanding cloud monomers have $\epsilon_{\mathbf{k},\text{trap}}\in[-8J,8J]$.  
Thus, only monomers with energies in the center of the band,
$-4J\leq\epsilon_{\mathbf{k},\text{trap}}\leq4J$, are able to escape 
into the rapidly expanding cloud because their energies match those of bosons in the vacuum.
This leads to four relations that define the momenta of
bosons that can leave the slowly expanding cloud,	
\begin{eqnarray}
k_y \geq \pm\arccos (\pm1-\cos k_x), \nonumber \\ 
k_y \leq \pm\arccos (\mp1-\cos k_x).
\label{expreg}
\end{eqnarray}
Obviously, this can only (if at all) approximate what happens in a real system
where interactions are finite and many-body effects are expected to affect the 
dynamics. In what follows, we look for signatures of the scenario above in the 
dynamics of the momenta occupations.
  
\begin{figure}[!t]
\centering
\includegraphics[width=0.35\textwidth]{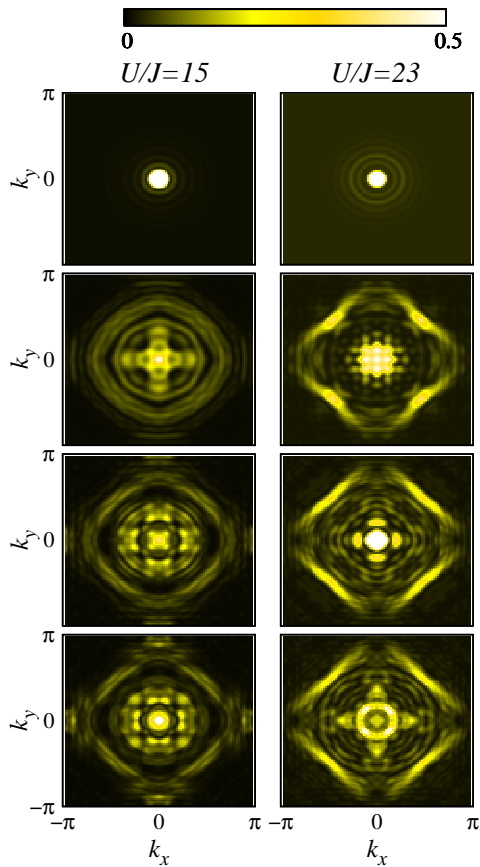}
\caption{(Color online) Momentum distribution functions for the cases for which 
density profiles are depicted in Fig.~\ref{fig:densityProfiles}.}
\label{fig:fullMomentum}
\end{figure}

In Fig.~\ref{fig:fullMomentum}, we show the momentum distribution function for the 
same systems and times depicted in Fig.~\ref{fig:densityProfiles}. While they exhibit
a rich structure, there are two specific features that are worth highlighting. The 
first one is that, at all times, there is a high population of modes close to 
$k_x=k_y=0$, i.e., there is a sizable fraction of bosons that either do not 
move or do so very slowly. 
The second one is that, during the expansion and with increasing $U/J$, 
a high population develops in modes with $k_y=\pi \pm k_x$ and $k_y=-\pi \pm k_x$. 
The latter ones are the modes that become most highly populated during the expansion 
of a Mott insulator with $n=1$ in the same geometry considered here 
\cite{jreissaty_carrasquilla_11}. They account for the fastest moving particles along 
the diagonals. This already suggests that different features in the momentum distribution 
may be associated to the slowly and rapidly expanding clouds seen in the density 
profiles in Fig.~\ref{fig:densityProfiles}, which we now study.

\begin{figure}[!t]
\centering
\includegraphics[width=0.34\textwidth]{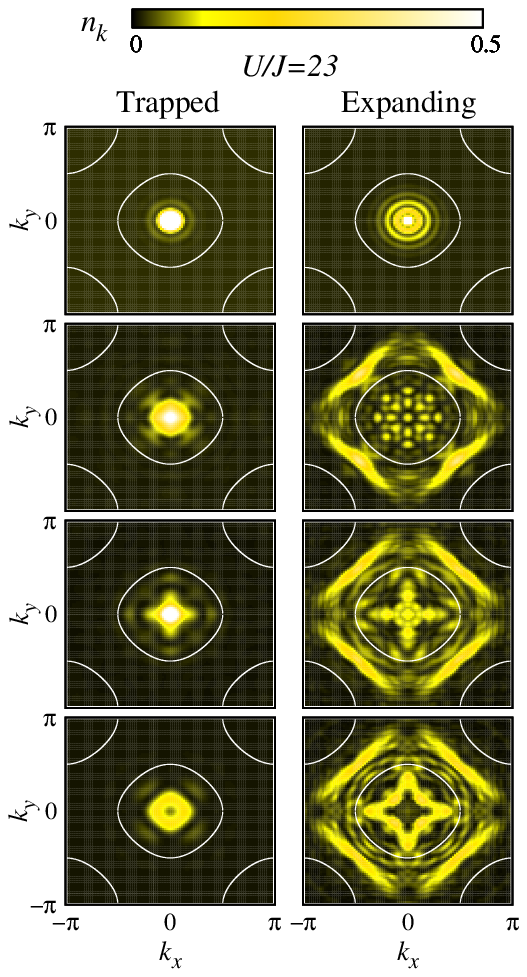}
\caption{(Color online) Momentum distribution of the slowly (left) 
and rapidly (right) expanding clouds for $U/J=23$, and the same times depicted in 
Figs.~\ref{fig:densityProfiles} and \ref{fig:fullMomentum}. White curves demarcate 
regions in momentum space for which monomers can or cannot escape the slowly expanding cloud 
(see text).}
\label{fig:U23_Momentum}
\end{figure}

Figure~\ref{fig:U23_Momentum} shows the momentum distribution \cite{supmat} 
of bosons that are part of the slowly (left) and rapidly (right) 
expanding clouds for $U/J=23$. In our calculations, 
as a slowly expanding cloud we take the bosons that remain in the region where at $t=0$ 
the system had $n>1$. (We find this to be a reasonable working definition despite 
the fact that, as seen in Fig.~\ref{fig:densityProfiles}, the region with 
$n>1$ does shrink during the expansion.) For $U/J=23$, the slowly expanding cloud 
consists of bosons that remain in a circle of radius $r_{\text{slow}}=8.5$ (in units of the 
lattice spacing) at the center of the system. The rest of the bosons were 
taken to be the rapidly expanding cloud. The right panels in Fig.~\ref{fig:U23_Momentum} 
show that the momentum distribution of the rapidly expanding cloud develops a 
diamondlike structure with a small population of the modes around $\mathbf{k}=0$. 
This structure becomes better defined as $U/J$ increases and is similar to the one 
observed during the expansion of a Mott insulator \cite{jreissaty_carrasquilla_11}.

More interestingly, the left panels in Fig.~\ref{fig:U23_Momentum} show that,
in addition to an almost uniformly populated background, particles in the slowly expanding  
cloud predominantly occupy low-momentum modes. Those modes belong to the $k$-space
region for which monomers are not allowed to escape \eqref{expreg}. This confirms
the relevance of the scenario proposed. During the dynamics, one can also see that 
the population of the momentum modes changes due to interactions, which allow some 
monomers to fall in the $k$ region where they can escape. Interactions are also 
important in the rapidly expanding cloud, where redistribution of momenta also 
occurs (right panels in Fig.~\ref{fig:U23_Momentum}). Overall, and similarly to results 
found in Refs.~\cite{rodriguez_manmana_06,jreissaty_carrasquilla_11,muth_petrosyan_12}, 
it is remarkable that the scenario previously discussed, which was developed 
for $U\rightarrow\infty$, provides a good qualitative understanding of the expansion 
of Bose-Hubbard systems with finite values of $U$ (greater than the bandwidth). 

\begin{figure}[!t]
\centering
\includegraphics[width=0.40\textwidth]{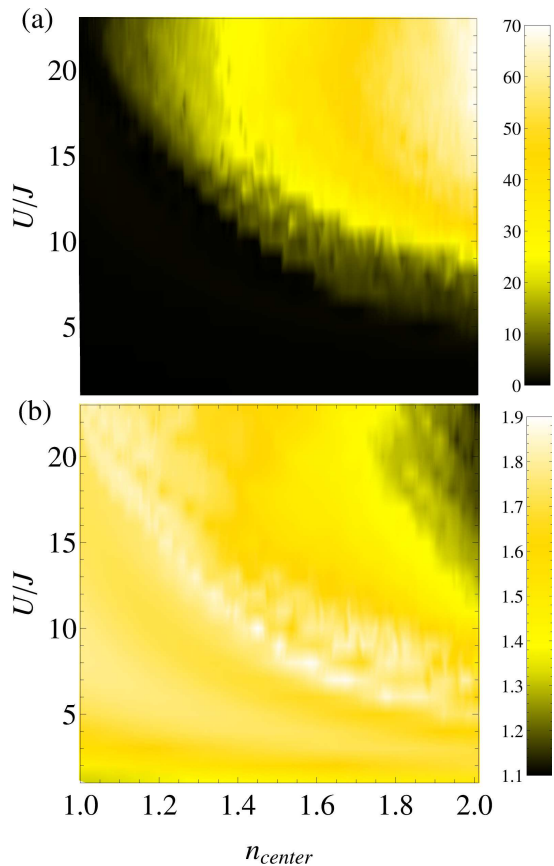}
\caption{(Color online) (a) Percentage of particles $P$ 
that remain in the slowly expanding region at times $t\simeq 30\hbar/J$, at which the 
fastest escaping particles have started reaching the edges of the $100\times100$ 
lattice utilized in our calculations \cite{supmat}, and (b) average radial 
expansion velocity $\bar{v}_r$ at times $t\simeq 15\hbar/J$, at which it is almost 
time independent \cite{supmat}, versus $n_{\text{center}}$ and $U/J$. For all 
systems, $V/J = 0.161$. $\bar{v}_r$ is given in units of lattice spacing 
times $J/\hbar$.}
\label{fig:Percent_Vel}
\end{figure}

We have studied the expansion of systems with values of $U/J$ between $1$ and $23$, and 
initial site occupancies in the center of the trap between $1.0$ and $2.0$, and have found 
that the self-trapping phenomenon discussed before occurs in an extended domain of 
interactions and fillings. Summarizing our results, 
Fig.~\ref{fig:Percent_Vel}(a) depicts the ratio $P$ between the time-averaged 
number of particles that remain in the slowly expanding cloud at times $t\simeq 30\hbar/J$ (longest
time simulated as particles reach the edges of our lattice) and the number of particles 
that were there initially vs $U/J$ and $n_{\text{center}}$ 
at $t=0$. At those long times $t$, $P$ exhibits a very weak time dependence.

Figure~\ref{fig:Percent_Vel}(a) shows that, for weak interactions ($U/J<5$), there is 
essentially no slowly expanding cloud for all initial densities $n_{\text{center}}$ considered. 
Under such conditions, doublons are not well-defined entities and self-trapping does not
occur. As $U/J$ increases beyond the bandwidth, self-trapping starts
to occur at lower occupancies in the center of the trap. This is expected as doublons 
become more difficult to break and less of them will dissociate and abandon the central
region. Furthermore, the scenario presented for the trapping of monomers becomes more 
relevant as $U/J$ increases, which further increases the stability of the region 
with $n>1$.

In recent experiments \cite{schneider_hacke_12,ronzheimer_schreiber_13}, 
a quantity of much interest has been the core expansion velocity and its 
alternative measure, the radial expansion velocity
$v_r \left(t\right) = \left(d/dt\right)\sqrt{R^2(t)-R^2(0)}$, where $R^2\left(t\right) = 
\left(1/N\right)\sum_i \langle \hat{n}_i \left(t\right) \rangle r^{2}_i$ and $N$ 
is the total number of particles. We have calculated $v_r \left(t\right)$ for all 
cases for which results were presented in Fig.~\ref{fig:Percent_Vel}(a). 
In general, this quantity is found to exhibit a transient regime in the early 
stages of the expansion and stabilizes to values that do not depend strongly 
on time before particles start reaching the edges of the lattice \cite{supmat}. 
Figure~\ref{fig:Percent_Vel}(b) depicts our results for such a stable radial 
velocity. 

In Fig.~\ref{fig:Percent_Vel}(b) one can see that, in the weakly interacting 
regime, $v_r \left(t\right)$ is low and slowly increases as the occupancy 
in the center of the trap increases. However, as $U/J$ increases, 
this velocity exhibits a maximum and steadily decreases as $n_{\text{center}}$
increases further. The initial increase can be intuitively understood 
as more interaction energy is present in the initial state and that energy is 
transformed into kinetic energy during the expansion. On the other hand, when 
$U/J$ exceeds a certain $n_{\text{center}}$-dependent value, the self-trapping 
mechanism discussed before sets in and the velocity starts to decrease. 
As expected, the value of $U/J$ at which the latter happens decreases with increasing 
filling and correlates with the onset of trapping in Fig.~\ref{fig:Percent_Vel}(b).

Note that the radial expansion velocities starting from the {\it ground state} of the 
trapped system are found to be nonzero as long as one is not deep in the self-trapping 
regime. This is in contrast to the experimental results for two-dimensional 
lattice bosons in Ref.~\cite{ronzheimer_schreiber_13} where, {\it after a quench in 
the interaction strength}, it was found that the radial expansion velocities drop quickly 
to zero as one moves away from the noninteracting limit. An important question 
that remains to be addressed, either theoretically with unbiased approaches 
or experimentally in the absence of an interaction quench, is whether starting 
from the ground state of a trapped two-dimensional system one obtains zero or 
nonzero radial expansion velocities. Independently of the possible outcome of
those studies, the self-trapping effect studied here within the mean-field 
approximation is expected to be robust, and has already been explored 
in one dimension \cite{muth_petrosyan_12,ronzheimer_schreiber_13}. 
 
In summary, we have shown that the expansion of bosons in a two-dimensional
lattice, after turning off a trap, can lead to a dynamical separation 
of the gas into a slowly and rapidly expanding clouds. The onset of this phenomenon 
depends on the ratio $U/J$ and the number of double occupancies at $t=0$, 
and was related to the difficulty of breaking doublons in the presence of strong 
repulsive interactions. We have also shown that, in contrast to the fermionic case, 
Bose enhancement can lead to the trapping of monomers. Furthermore, we analyzed 
the effect that self-trapping has on the radial expansion velocity of the system, 
which could be potentially studied in experiments with ultracold bosons in optical 
lattices.

\begin{acknowledgments}
This work was supported by the U.S. Office of Naval Research. We are grateful to 
F. Heidrich-Meisner and U. Schneider for stimulating discussions.
\end{acknowledgments}

\onecolumngrid

%\vspace*{0.4cm}

\begin{center}

{\large \bf Supplementary Materials:
\\ Self trapping in the two-dimensional Bose-Hubbard model}\\

\vspace{0.6cm}

Andrew Jreissaty$^{1,2}$, Juan Carrasquilla$^{2,1}$, and Marcos Rigol$^2$\\
\ \\
$^1${\it Department of Physics, Georgetown University, Washington DC, 20057, USA}\\
$^2${\it Department of Physics, The Pennsylvania State University, University Park, Pennsylvania 16802, USA}\\

\end{center}

\vspace{0.6cm}

\twocolumngrid

\subsection{Gutzwiller mean-field approximation}

The results presented in the main text were obtained using the Gutzwiller 
mean-field approximation, discussed in detail in Refs.~\cite{amico_penna_98,jaksch_venturi_02,
jreissaty_carrasquilla_11,snoek_11}. Within this approach, the ground state 
and time-evolving states are approximated by a product of individual site wavefunctions, 
each of which is a linear combination of Fock states:
\begin{equation}
\vert \psi_{\text{MF}} \rangle = \prod_{i=1}^L \sum_{n=0}^{n_c} \alpha_{in} \vert n \rangle_i
\label{mf}
\end{equation}
Here, $L=l\times l$ is the total number of lattice sites and $l$ is the linear size, $|n\rangle_i$ 
is a Fock state corresponding to $n$ particles at site $i$, $n_c$ is the cutoff determining the 
maximum number of particles allowed per site ($n_c=5$ in our study), and $\alpha_{in}$ are 
the coefficients associated with each Fock state. The ground state of the system is 
obtained by finding the coefficients $\alpha_{in}$ that minimize the mean-field energy. 
The equations for the time evolution of the coefficients $\alpha_{in}$ during the expansion 
are obtained utilizing the time-dependent variational principle \cite{amico_penna_98,
jaksch_venturi_02,jreissaty_carrasquilla_11,snoek_11}, and we integrate 
them numerically using a fourth-order Runge-Kutta with a discretization time $\delta t=2.0 \times 10^{-4}$.

\subsection{Momentum distributions for $U/J=15$}

\begin{figure}[!t]
\centering
\includegraphics[width=0.35\textwidth]{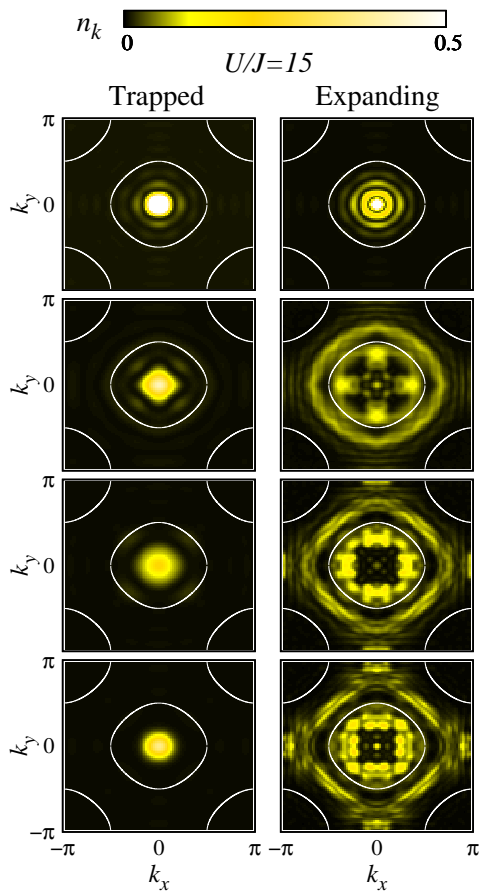}
\caption{(Color online) Momentum distribution function of the slowly (left panels) 
and rapidly (right panels) expanding clouds for $U/J=15$, and the same times depicted in 
Figs.~1 and 2 in the main text. White curves demarcate regions in momentum space for 
which monomers can or cannot escape the slowly expanding cloud (see main text).}
\label{fig:U15_Momentum}
\end{figure}

Figure~\ref{fig:U15_Momentum} shows the time evolution of the momentum distribution 
function of the slowly and rapidly expanding clouds in a system with $U/J=15$. The 
momentum distribution function is computed by Fourier transforming the one-particle density 
matrix as
\begin{equation}
n^{\beta}_{\mathbf k}\left(t\right)=\frac{1}{L}\sum_{i,j}e^{\mathbf{k}\left(\mathbf{r}_i-\mathbf{r}_j \right)} 
\langle \psi^{\beta}_{\text{MF}} \vert b^{\dag}_ib_j\vert \psi^{\beta}_{\text{MF}} \rangle,
\end{equation}
where the sums go over the entire lattice, $r_i$ is the position of the $i$-th site in the lattice, 
and $\beta=s,r$ indicates whether we are dealing with the slowly or rapidly expanding 
cloud, respectively. The mean-field states $\vert \psi^{\beta}_{\text{MF}} \rangle$, written in 
terms of new sets $\alpha^{\beta}_{in}$, are defined through the full set of coefficients 
$\alpha_{in}$ in Eq.~\eqref{mf} and the radius of the circle defining the slowly expanding cloud 
$r_{\text{slow}}$. For the slowly expanding region they are 
defined as
\begin{equation*}
\begin{array}{rcl}
\alpha^{s}_{in}= &  \alpha_{in} & \text{if } |\mathbf{r}_i| \le r_{\text{slow}},\\
\alpha^{s}_{i0}= & 1  & \text{if } |\mathbf{r}_i| > r_{\text{slow}},\\
\alpha^{s}_{i n \neq 0}= & 0 & \text{if } |\mathbf{r}_i| > r_{\text{slow}},
\end{array} 
\end{equation*}
while for the rapidly expanding region they are defined as
\begin{equation*}
\begin{array}{rcl}
\alpha^{r}_{in}= &  \alpha_{in} & \text{if } |\mathbf{r}_i| \ge r_{\text{slow}},\\
\alpha^{r}_{i0}= & 1  & \text{if } |\mathbf{r}_i| < r_{\text{slow}},\\
\alpha^{r}_{i n \neq 0}= & 0 & \text{if } |\mathbf{r}_i| < r_{\text{slow}}.
\end{array} 
\end{equation*}
The results for the time evolution of the density profiles and full momentum distributions of this
system were shown in Figs.~1 and 2 in the main text. We note that at $t=0$, and later 
times, the occupation of some momentum modes may exceed the scale depicted in 
Fig.~\ref{fig:U15_Momentum} and in the figures in the main text. We have truncated
those occupancies at $0.5$ so that the most important features in the momentum distribution 
function can be discerned in the plots.

The results depicted in Fig.~\ref{fig:U15_Momentum} are qualitatively similar 
to those for $U/J=23$ in Fig.~3 in the main text. The most apparent quantitative difference 
is that the diamond-like structures in the expanding cloud are significantly more populated 
for $U/J=23$ than for $U/J=15$. This is expected from the analysis in 
Ref.~\cite{jreissaty_carrasquilla_11}.

\subsection{Radial expansion velocity}

\begin{figure}[!h]
\centering
\vspace{0.7cm}
\includegraphics[width=0.47\textwidth]{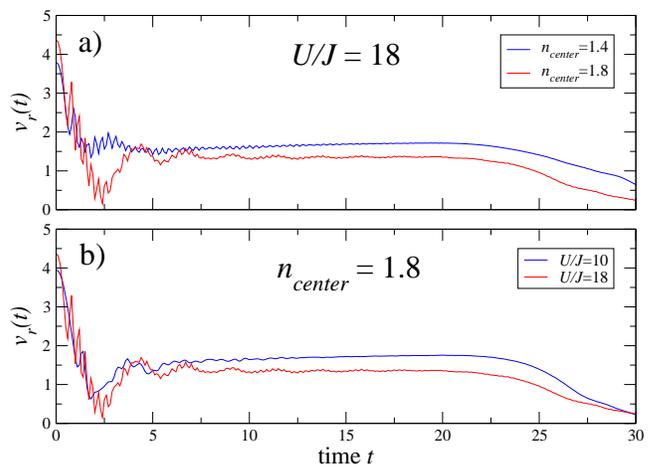}
\caption{(Color online) Time evolution of the radial expansion velocity $v_r(t)$ for (a) two 
systems with $U/J=18$, and $n_{\text{center}}=1.4$ and $n_{\text{center}}=1.8$, and (b) two 
systems with $n_{\text{center}} = 1.8$, and $U/J=10$ and $U/J=18$. The time is reported in 
units of $\hbar/J$ and velocity in units of lattice spacing times $J/\hbar$}
\label{fig:VelocityFigure}
\end{figure}

Figure~\ref{fig:VelocityFigure} shows the time evolution of the radial expansion velocity 
after turning off the trap for $U/J=18$  and two different values of $n_{\text{center}}$ 
[Fig.~\ref{fig:VelocityFigure}(a)] and for $n_{\text{center}}=1.8$ and two different 
values of $U/J$ [Fig.~\ref{fig:VelocityFigure}(b)]. One can see there that, after a 
transient regime, the velocities become almost time independent before dropping abruptly
when particles start reaching the edges of the lattice.

For fixed $U/J = 18$, Fig.~\ref{fig:VelocityFigure}(a) shows that an increase in 
$n_{\text{center}}$ from $1.4$ to $1.8$ results in a decrease in the radial expansion velocity. 
This is understandable as the initial number of doublons increases, which enhances self trapping. 
For fixed $n_{\text{center}} = 1.8$, Fig.~\ref{fig:VelocityFigure}(b) shows that an increase 
in $U/J$ from $10$ to $18$ also causes $v_r(t)$ to decrease. This is because doublons 
become more difficult to break and self trapping is also enhanced. These two effects were 
discussed in the main text in the context of the results presented in Fig. 4(b).\\

\subsection{Percentage of slowly expanding particles and asymptotic radial expansion velocity}

\begin{figure}[!h]  
\centering
\includegraphics[width=0.42\textwidth,angle=-90]{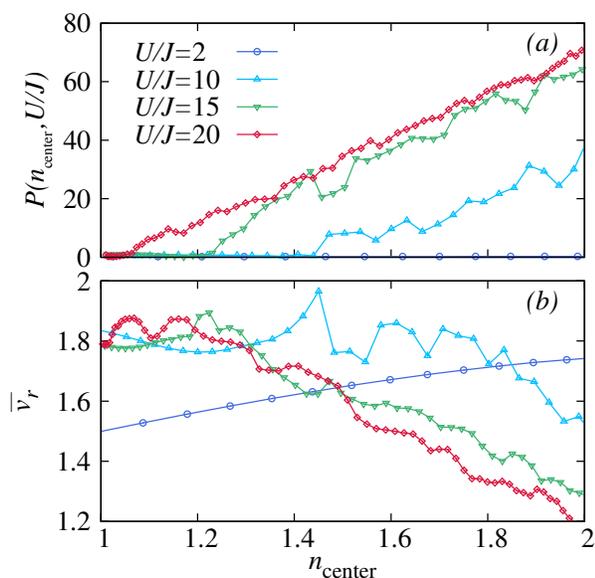}
\caption{(Color online) (a) Percentage of particles $P$ 
that remain in the slowly expanding region at times $t\simeq 30\hbar/J$, as function of 
$n_{\text{center}}$ for different values of $U/J$. (b) Average radial 
expansion velocity $\bar{v}_r$ at times $t\simeq 15\hbar/J$. The average velocity 
is given in units of lattice spacing times $J/\hbar$}
\label{fig:Percentage_and_velocity}
\end{figure}

Figure~\ref{fig:Percentage_and_velocity}(a) shows the evolution of the percentage of particles 
$P$ initially confined that remain in the slowly expanding region at times $t\simeq 30\hbar/J$
as a function of $n_{\text{center}}$ and for several values of $U/J$. $P$ is computed as a time 
average over 15 time steps, from $t=28.6\hbar/J$ to $t=30.0\hbar/J$ (every $t=0.1\hbar/J$). 
For small values of $U/J<2$ there are no particles that remain in the slowly expanding region 
for any initial occupancy at the center of the lattice, whereas for larger values of $U/J$
there is a finite amount of particles that remain in the slowly expanding cloud for large 
enough $n_{\text{center}}$. The value of $n_{\text{center}}$ at which the self-trapping 
sets in decreases as the interaction $U/J$ is increased, as expected. 

Figure~\ref{fig:Percentage_and_velocity}(b) shows the time-averaged 
radial expansion velocity $\bar{v}_r$ as a function of $n_{\text{center}}$ and for 
several values of $U/J$. The time average is computed over 20 time steps starting from 
$t=14.9\hbar/J$ to $t=16.8\hbar/J$ (every $0.1\hbar/J$). For small $U/J$, 
$\bar{v}_r$ slowly increases because the increased interaction energy that consequently 
transforms into a higher kinetic energy upon expansion. For large values of $U/J$, 
$\bar{v}_r$ first increases and then decreases after reaching a maximum as 
$n_{\text{center}}$ increases. This effect can be understood as 
the system converts all its interaction energy (which increases with increasing $U/J$) 
into kinetic energy whenever no self-trapping occurs. After self-trapping sets in, 
which occurs for smaller values of $n_{\text{center}}$ as $U/J$ increases,
the radial velocity starts to decrease because an increasingly fraction of particles 
remain in the slowly expanding region at the center of the system.\\

\subsection{Radially averaged densities after a long expansion time}

\begin{figure}[!h]
\centering
\includegraphics[width=0.32\textwidth,angle=-90]{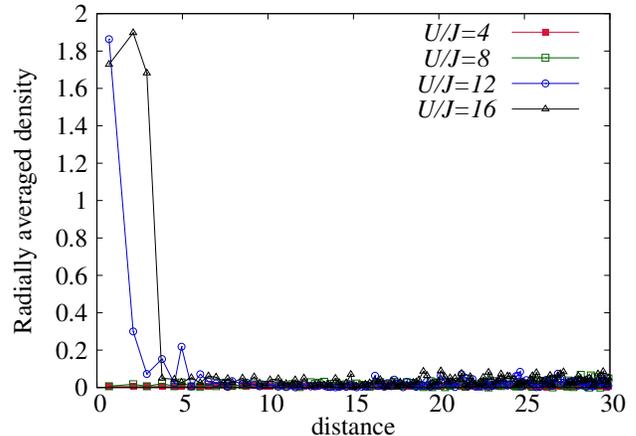}
\caption{(Color online) Radially averaged density profiles after $t=20\hbar/J$ for different values
of $U/J$. The distance is given in units of the lattice spacing. The initial site occupancy at the 
center of the trap was taken to be $n_c=1.5$ and the strength of the trapping potential to be $V/J=0.161$.}
\label{fig:radiallyaveraged}
\end{figure}

Figure~\ref{fig:radiallyaveraged} depicts the radially averaged density profiles
after $t=20\hbar/J$ for several values of $U/J$ as a function of the distance from the center
of the system. Notice that the density profiles in this study do not possess radial symmetry and
some information is lost by taking the radial average. Nevertheless, such information can be 
contrasted directly with experiments and helps to detect the onset of self-trapping. 

Figure~\ref{fig:radiallyaveraged} shows that, for small values of $U/J=4,8$, there is no 
self-trapping in the time scales considered. For $U/J=12,16$, on the other hand, self-trapping
does occur. The data in this figure are consistent with Fig.~4 of the main text, for which, 
at $n_c=1.5$, self-trapping becomes apparent at $U/J\simeq10$.

%

%\bibliography{references}

\end{document}